\documentclass{aastex}
\usepackage{spr-astr-addons}
\usepackage{epstopdf}
\begin{document}
\title{Two phase formation of massive elliptical galaxies: study through cross-correlation including spatial effect}
\shorttitle{Two phase formation of ETGs}
\shortauthors{Modak et al. 2017}
\author{Soumita Modak\altaffilmark{1}}
\and
\author{Tanuka Chattopadhyay\altaffilmark{2}}
\and
\author{Asis Kumar Chattopadhyay\altaffilmark{1}}
\altaffiltext{1}{Department of Statistics, University of Calcutta, 35 B.C. Road, Kolkata 700019, India.}
\altaffiltext{2}{Department of Applied Mathematics, University of Calcutta, 92 A.P.C Road, Kolkata 700009, India.}

\begin{abstract}
Area of study is the formation mechanism of the present-day population of elliptical
galaxies, in the context of hierarchical
cosmological models accompanied by accretion and minor mergers. The present work investigates the formation and evolution of several
components of the nearby massive early-type galaxies (ETGs) through cross-correlation function
(CCF), using the spatial parameters right ascension ($RA$) and
declination ($DEC$), and the intrinsic parameters mass ($M_*$) and size. According to the astrophysical terminology, here these variables, namely mass, size, $RA$ and $DEC$ are termed as parameters, whereas the unknown constants involved in the kernel function are called hyperparameters. Throughout this paper, the parameter size is used to represent the effective radius ($R_e$). Following \citet{hua13a}, each nearby ETG is divided into three parts on the basis of its $R_e$ value. We study the CCF between each of these three components of nearby massive ETGs and the ETGs in the high
redshift range, $0.5< z\leq2.7$. It is found that the innermost
components of nearby ETGs are highly correlated with ETGs in the
redshift range, $2< z\leq2.7$, known as `red nuggets'. The
intermediate and the outermost parts have moderate correlations with
ETGs in the redshift range, $0.5< z\leq0.75$. The quantitative
measures are highly consistent with the two phase formation
scenario of nearby massive ETGs, as suggested by
various authors, and resolve the conflict raised in a previous work \citep{de14} suggesting other possibilities for the formation of the outermost part. A probable cause of this improvement is the inclusion of the spatial effects in addition to the other
parameters in the study.
\end{abstract}


\section{Introduction}

\cite{hub34} first studied the frequency distribution of
galaxies according to their positions (i.e. $RA$ and $DEC$) in the universe and the distribution of pairwise distances was
found to be strongly skewed, which indicates that the galaxies have
spatial clustering nature. The clustering nature is intrinsic
compared to uniform distribution \citep{bok34,mow38}.
Numerous studies have been carried out to explore the clustering
nature using position coordinates (viz. $RA$ and $DEC$) \citep{cha52,zwi33,lim53,lim54,ney54}. \citet{de14}
investigated the clustering nature with respect to the intrinsic
properties of the galaxies through mass and size.

The discovery of compact massive ETGs at high
redshift has raised questions on the plausible formation
mechanisms for the present-day population of elliptical galaxies from the classical view of a single event, e.g. monolithic
collapse \citep{lar75,car84,ari87} or a major merger \citep{too72,ash92,zep00,ber11,pri13}. Instead, the single events have to be embedded in
hierarchical merging, called minor mergers accompanied by
continuous accretion \citep{for97,jeo07,mon08,kav11,blu12,new12,sha12,sha17}. Interested readers may also consult the references listed in these papers. Recent discovery of
compact high redshift galaxies \citep{dad05,cap09,dam11,ono12} and intermediate redshift galaxies, having stellar masses
and sizes increased by a factor of $3-4$ \citep{van10,pap12,szo12},
suggests two phase formation of nearby massive
ETGs \citep{ose10,for11,gob11,van15,del16,vul16}.
First, an intense dissipational process like accretion \citep{dek09} or major merger forms an initially compact
inner part, then a slower phase starts when the outermost part
is developed through non-dissipational process, e.g. dry merger.
This two phase model \citep{ose10,joh12} severely challenges the classical `single
event' model. \citet{hua13b} explored the two phase theory through matching `median' values of
the two systems, which suffers from the following limitations. They have considered only
univariate data like `mass' or `size' at a time. The present study shows that the ETGs in the redshift range, $0.5< z\leq2.7$, have estimated correlation (Pearson product-moment correlation coefficient) between mass and size as $r_p=0.348$, with $p-$value $\simeq0$ for testing the null hypothesis $H_0:$ correlation between mass and size is zero against the alternative $H_1:$ correlation between mass and size is non-zero. Three components of nearby ETGs (defined in Section $2$) have estimated correlations as $r_p=0.817$, $r_p=0.623$ and
$r_p=0.668$ respectively, and each has $p-$value $\simeq0$ for the above test. It indicates that mass and size of the ETGs are correlated and hence the use of univariate median matching is not suitable for them. Also, median causes loss of information, as it does not directly use all the observations of a given data set. But, bivariate study of mass and size through CCF
 \citep{de14} suggested
three phase formation of nearby massive ETGs.
To solve this conflict, we study the CCF by combining the intrinsic parameters mass and size with the spatial effects $RA$ and $DEC$. As mass-size bivariate data are
significant for exploring intrinsic clustering nature and spatial
coordinates are significant for spatial clustering, we combine the above two to study the interaction
between them.

The present work computes the CCF between the different
components of nearby massive ETGs and the high
redshift $(0.5< z\leq2.7)$ ETGs. To find the distances between the ETGs, we consider the Euclidean norm, which is suitable for linear parameters like mass and size, but cannot be used for angular parameters like $RA$ and $DEC$. As our study includes both types of parameters, we consider the Euclidean metric based on mass, size, linearized $RA$ (say, $RA_l$) and linearized $DEC$ (say, $DEC_l$), in which values of $RA$ and $DEC$ are linearized throughout the analysis. Here, due to noisy or sparsely distributed ETGs, the raw data cannot be used in studying the CCF. Hence, we compute the CCF, using the relevant information in terms of the kernel principal components (KPCs). Here the KPCs are extracted from the ETGs through kernel principal component analysis (KPCA) with the help of the positive definite kernel proposed in \citet{mod17}. It gives the result which is highly consistent with the two phase formation scenario of nearby massive ETGs. It also resolves the conflict raised in a previous work \citep{de14} suggesting other possibilities for formation of the outermost part of nearby massive ETGs. The paper is organized as follows.
In Section $2$, we discuss the data sets whereas Section $3$ describes the
methods. The results and discussion are given in
Section $4$.

\section{Data sets}

In \citet{hua13a}, a nearby ETG is divided into three parts defined as the inner
component with $R_e$ $\leq1$ kpc, the intermediate
component with $R_e\sim2.5$ kpc, and the outermost
component with $R_e\sim10$ kpc. Let the effective radii and the masses for these three parts be denoted by $R_{e,1},R_{e,2},R_{e,3}$ and $M_{*,1},M_{*,2},M_{*,3}$ respectively. Here each galaxy has unique values for $RA$ and $DEC$ (regardless of its different parts), which are transformed from angular form to linear form (see, Subsection 3.1, for details of the transformation method) for our study. Thus, we have data for $70$ nearby massive ($M_*>10^8 M_\odot$) ETGs on eight parameters, namely $R_{e,1}$, $R_{e,2}$, $R_{e,3}$, $M_{*,1}$, $M_{*,2}$, $M_{*,3}$, $RA_l$ and $DEC_l$. Now, to include the spatial effect in the study, we first cluster these $70$ ETGs, based on the above-mentioned eight parameters, into three clusters. Here, for clustering we use the k-medoids clustering method based on the Euclidean distance \citep{kau90}. It results in the first cluster with the minimum value for $<R_{e,1}>$ ($<R_{e,1}>=$ arithmetic mean of observations on $R_{e,1}$), the second cluster with medium value for $<R_{e,2}>$ and the third cluster with the maximum value for $<R_{e,3}>$.
Then, we choose $R_{e,1}$, $M_{*,1}$, $RA_l$, $DEC_l$ values for the ETGs from the first cluster denoted as data set $1$, which represents the innermost part of the nearby ETGs. Next, we consider $R_{e,2}$, $M_{*,2}$, $RA_l$, $DEC_l$ observations for the second cluster which form data set $2$, representing the intermediate part. Lastly, data set $3$ includes $R_{e,3}$, $M_{*,3}$, $RA_l$, $DEC_l$ values from the third cluster which denotes the sample from the outermost part. Thus, data sets $1-3$ of sizes $15, 25$ and $30$ respectively, have observations on four parameters mass, size, $RA_l$ and $DEC_l$. Depending on the division of $R{e,i}$ $(i=1,2,3)$ value in three groups, there may be $27$ different combinations as each $R{e,i}$ $(i=1,2,3)$ can be chosen in $3$ ways, although some combinations might not give rise to significant variations in the CCF. In order to check for the robustness
of our results which depend upon
the chosen representative samples, we have also considered
other distinct choices for data sets $1-3$ and obtained similar results.

For comparison, we collect data on mass, size, $RA_l, DEC_l$ for $1012$ ETGs in the high redshift range (viz. $0.5< z\leq2.7)$ from the following sources.
$364$ ETGs $(0.50< z \leq 2.67)$ from \citet{dam11},
$248$ ETGs $(1.49\leq z \leq 1.79)$ from \citet{pap12},
$142$ ETGs $(0.70\leq z \leq 1.31)$ from \citet{che13},
$177$ ETGs $(0.52\leq z \leq 2.35)$ from \citet{szo13} and
$81$ ETGs $(1.28\leq z \leq 1.50)$ from \citet{mcl13}.
These data sets $4-8$ of respective sizes $168, 218, 206, 391$ and $29$, contain massive ellipticals (with $M_*>
10^8 M_\odot$) in the redshift zones (defined similarly as in De et al. $(2014)$) $0.5<z\leq 0.75$,
$0.75<z\leq 1$, $1<z\leq 1.4$, $1.4<z\leq 2$ and $2<z\leq 2.7$ respectively.

\section{Method}

\subsection{Linearization of angular data}

\begin{figure}
\centering
\includegraphics[width=1\textwidth]{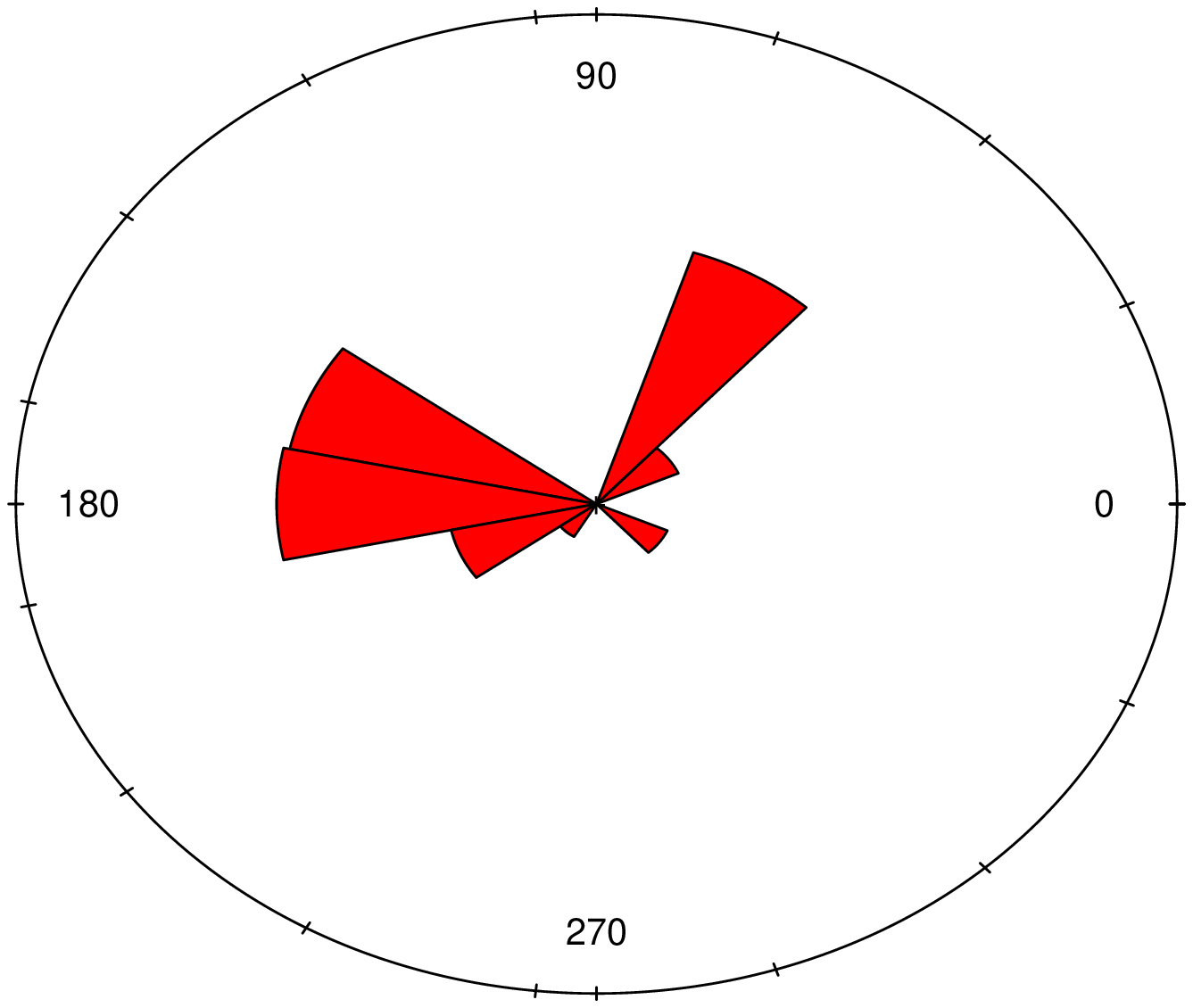}
\caption{Histogram of right ascension ($RA$) for the ETGs with $0.5\leq z\leq0.75$ (data set $4$).}\label{RA}
\end{figure}

\begin{figure}
\centering
\includegraphics[width=1\textwidth]{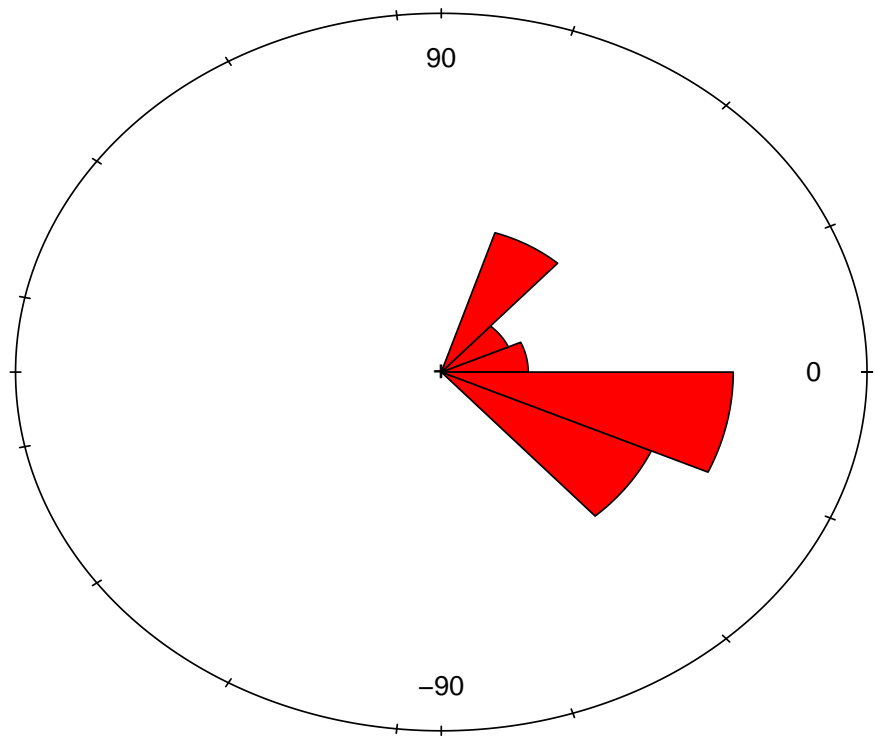}
\caption{Histogram of declination ($DEC$) for the ETGs with $0.5\leq z\leq0.75$ (data set $4$).}\label{Dec}
\end{figure}

As we consider the Euclidean distance in estimating the CCF, we transform the angular data on $RA$ and $DEC$ into some linear form by the help of the method proposed in \citet{cha15}. Let $\theta$ be the angular data
with the unique mode $\phi$, then its linear form is given by $1-cos(\theta-\phi)$. If $\theta$ has two
modes $\phi_1$ and $\phi_2$, then its linear form is
given by $max\{1-cos(\theta-\phi_1),1-cos(\theta-\phi_2)\}$. Now, mode of $\theta$ can be estimated
from its
histogram, in which the circumference of the circle
is split into groups specified by bins and the radii of the
corresponding sectors are computed as equal to the square root of
the relative frequencies of observations in each group. For example, see Fig. \ref{RA}, the histogram of $RA$ for the ETGs in the redshift zone $0.5<z\leq 0.75$, i.e. data set $4$, indicates a bimodal distribution with $\phi_1
\sim 45^{\circ}$ and $\phi_2 \sim 180^{\circ}$. Similarly, the histogram of $DEC$ for data set $4$ (Fig. \ref{Dec}) indicates its unimodal density with $\phi \sim 0^{\circ}$.

\subsection{Compatibility test}

Since the data sets $4-8$ have
different sources, they obviously suffer from selection biases and
errors. Hence, compatibility test is necessary before combining them into a study. So, we
perform the Duda-Hart test \citep{dud73}, for
each of the four parameters, to check homogeneity between two data
sets from different sources
(Table \ref{table:t1}). The test is performed for data sets within the same
redshift zone, as the galaxies have undergone cosmological
evolution via merger and accretion \citep{kho06,del07,guo08,kor09,hop10,naa13}. Data from \citet{dam11} contains the maximum number of galaxies within the entire
redshift zone $(0.5< z\leq2.7)$. So, we perform the testing between
the above-mentioned data set and the other sets. It is clear
from Table \ref{table:t1} that all the tests are accepted at $5\%$
level of significance with high or moderate $p-$values. Hence, we
assume that the combined data sets are compatible with each other
with respect to the four parameters size, mass,
$RA_l$ and $DEC_l$.

\subsection{Completeness test}

For testing completeness of the combined data sets $4-8$, we
use the $V/V_{max}$ test. The test was first used by \citet{sch65}
for studying the space distribution of a complete sample of
radio quasars. Let $F_m$ be the limiting flux within maximum
distance $r_m$ and $r$ be the radial distance of a quasar. Let us also define $V(r)$ and $V_{max}$ as $V(r)=\frac{4\Pi r^3}{3}$ and $V_{max}=\frac{4\Pi r_m^3}{3}$. Then, if all the
quasars follow a uniform distribution over the entire range of
observation, $V/V_{max}$ will be uniformly distributed over
$[0,1]$ with $<V/V_{max}>=0.5$. We obtain
$<log|R_e|/log|R_{e,{max}}|>= 0.445$ (removing $65$ outliers), $<log|M|/log|M_{max}|>=
0.862$, $<logRA/logRA_{l,max}>=0.537$ and
$<logDEC/logDEC_{l,max}>=0.363$ (removing 145 outliers) (here, suffix `$max$' represents the
maximum value of the corresponding parameter). As the mean values for all the parameters, except mass, are close to $0.5$, we can conclude that
the combined data set is complete in $75\%$ (i.e. $3$ out of $4$) of its parameters.

\begin{figure}
\centering
\includegraphics[width=1\textwidth]{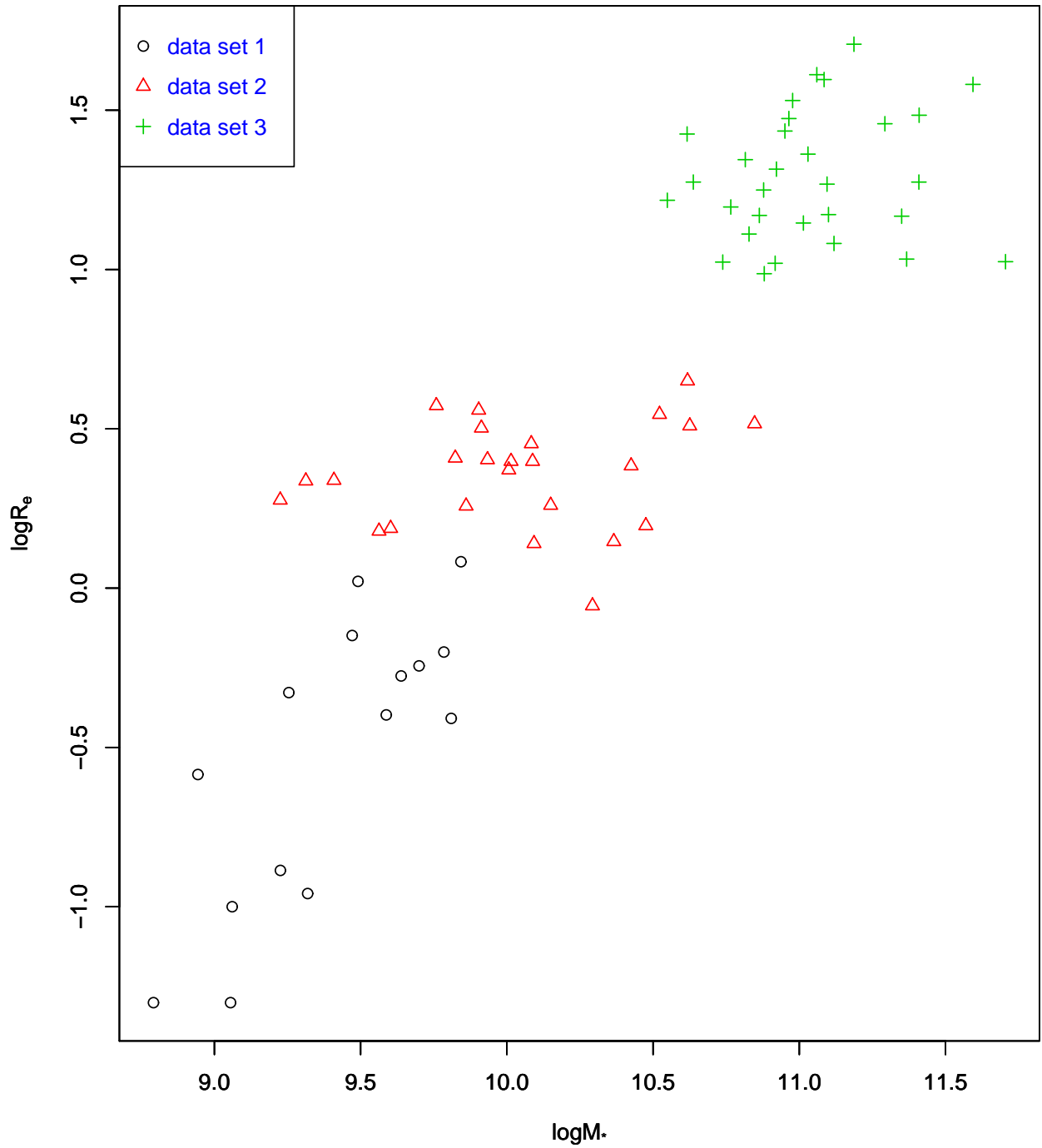}
\caption{Mass versus size plot for the representative samples of the
three components of nearby ETGs.}\label{MR1}
\end{figure}

\begin{figure}
\centering
\includegraphics[width=1\textwidth]{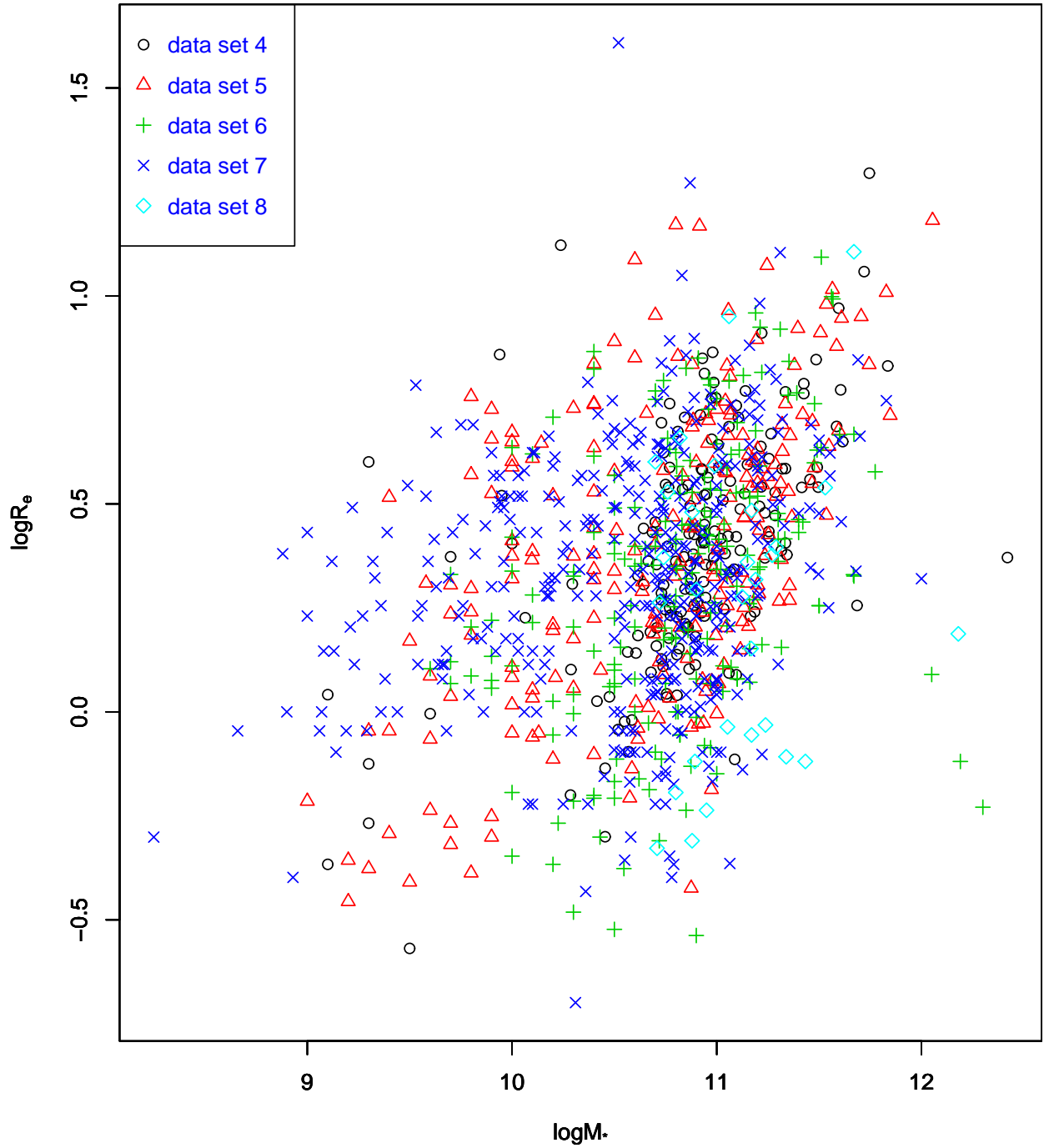}
\caption{Mass versus size plot for the ETGs in the high redshift zone.}\label{MR2}
\end{figure}

\begin{figure}
\centering
\includegraphics[width=1\textwidth]{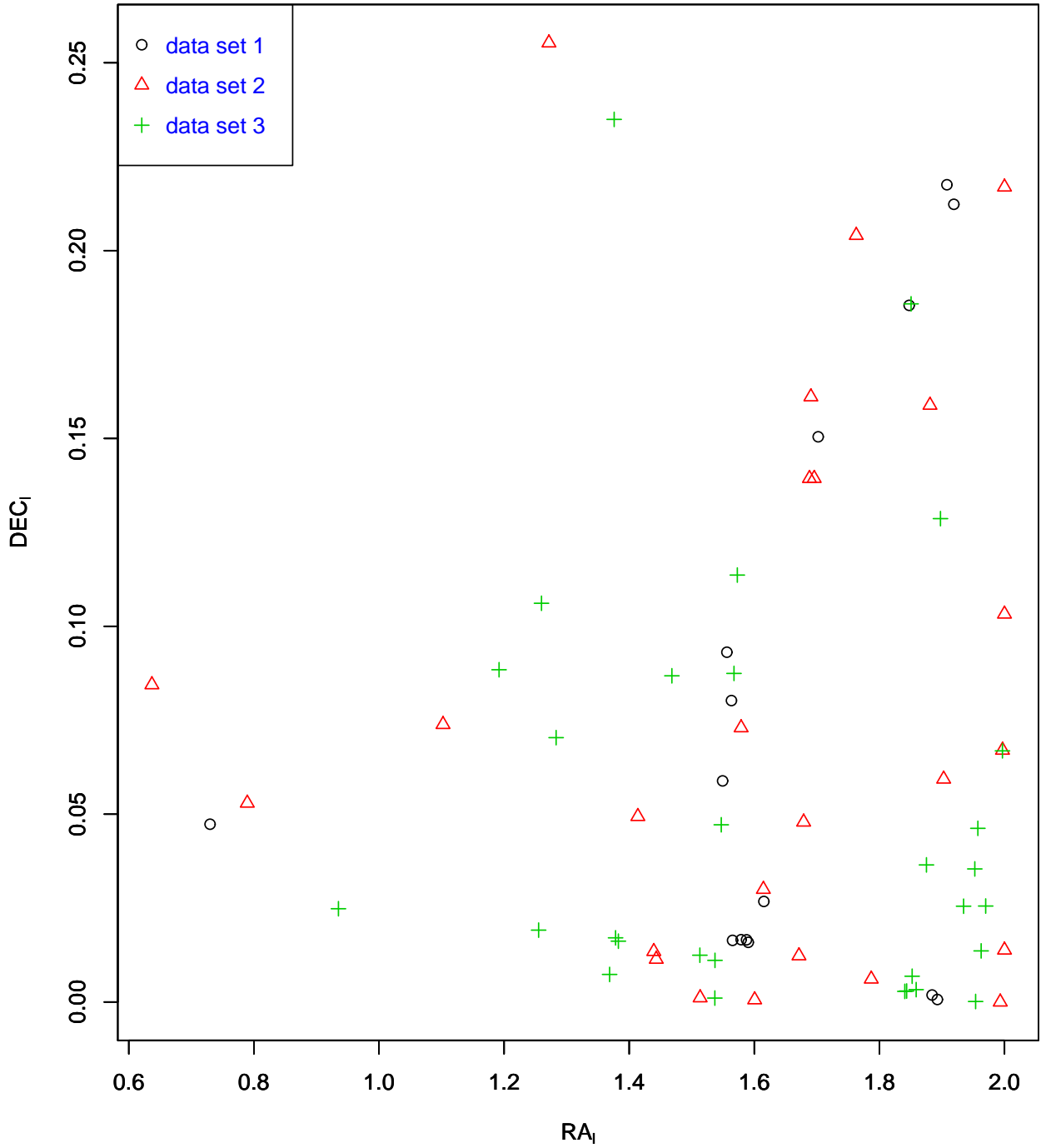}
\caption{Linearized $RA$ ($RA_l$) versus linearized $DEC$ ($DEC_l$) plot for the representative samples of the
three components of nearby ETGs.}\label{RD1}
\end{figure}

\begin{figure}
\centering
\includegraphics[width=1\textwidth]{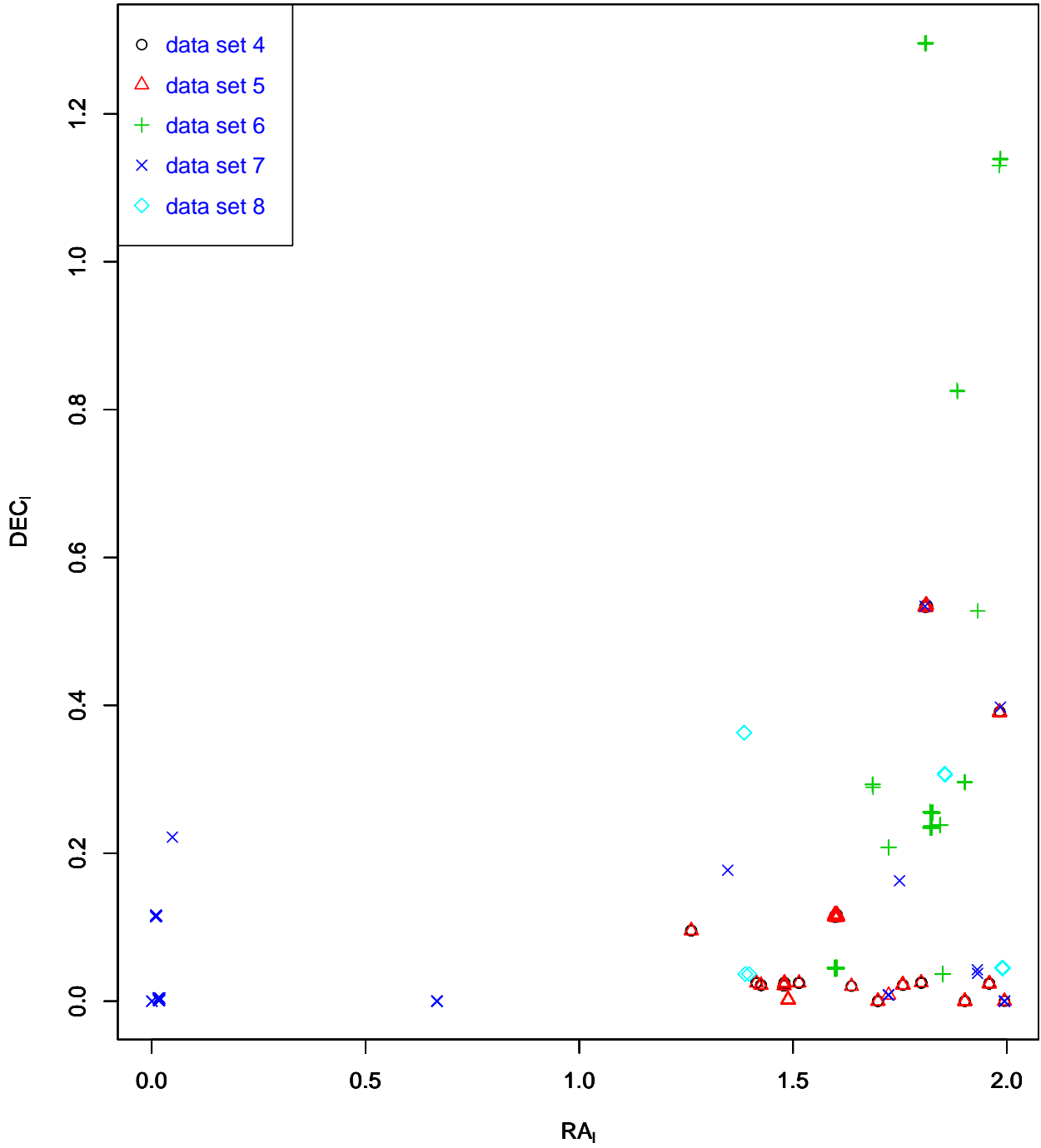}
\caption{$RA_l$ versus $DEC_l$ plot for the ETGs in the high redshift zone.}\label{RD2}
\end{figure}

This method involves only point estimation of the expectation of Uniform $(0,1)$ by its sample mean, which is always affected by outliers. Here, by outliers we mean the observations which are ``far away" from the average value of the remaining observations. Hence, a parallel nonparametric test, which is unaffected by outliers, is also preferred. So, we check completeness by performing the Kolmogorov-Smirnov test \citep{kol33,cor14} for the following testing problems with
$H_{01}$: Distribution of $logR_e/logR_{e,{max}}$ is Uniform over $[-1,1]$, $H_{02}$: Distribution of $logM/logM_{max}$ is Uniform over $[-1,1]$,  $H_{03}$: Distribution of $logRA_l/logRA_{l,max}$ is Uniform over $[0,1]$, and  $H_{04}$: Distribution of $logDEC/logDEC_{l,max}$ is Uniform over $[0,1]$, for each of which the alternative is ``the null hypothesis does not hold". The $p-$values for all the above tests came out to be almost zero, which proves that the combined data set is complete.
For comparison, we also plot the
combined data sets $1-3$ and $4-8$ in the mass-size plane (Figs.
$\ref{MR1}$ and $\ref{MR2}$ respectively) and in the $RA_l-DEC_l$ plane
(Figs. $\ref{RD1}$ and $\ref{RD2}$ respectively).

\subsection{Cross-correlation function (CCF)}

The theory of the CCF $(\xi)$ was first introduced by
\citet{ney52} for studying galaxy clustering on the
basis of four assumptions, namely $(i)$ galaxies have a natural
tendency to occur in groups, $(ii)$ the number of galaxies varies in each
group following certain probabilistic law, $(iii)$ the galaxies in
each group also follow certain probabilistic distribution, and $(iv)$
group centers in space follow quasi-uniform distribution.
Subsequently, the above theory has been discussed by several
authors, e.g. \citet{pee80,mar02,bla06,de14}, etc. We consider the following estimator of the CCF \citep{bla06} as,
\begin{equation}\label{ECCF}
\hat\xi(d)=\frac{D_1D_2(d)-D_1R_2(d)-D_2R_1(d)+R_1R_2(d)}{R_1R_2(d)},
\end{equation}
where $d$ denotes the distance computed between each of the unique pairs of galaxies in two data sets. $D_1D_2(d)$ denotes a pair count between the observed data sets $D_1$ and $D_2$, i.e. frequency corresponding to a bin with separations of $d\pm \delta d/2$ in the histogram of $d$. Let $R_i$ denotes the unclustered sample drawn from the data set $D_i,i=1,2.$ Then, $D_iR_j(d)$ denotes
pair count between data sets $D_i$ and $R_j$, $i,j=1,2$ but $i\neq j$, and $R_1R_2(d)$ denotes
pair count between data sets $R_1$ and $R_2$.

In our study, $\hat\xi$ is computed, based on mass, size, $RA_l$ and $DEC_l$, where $d$ is the normalized
Euclidean distance computed between each of the unique pairs of galaxies in two data sets.
But, due to noisy or sparsely distributed data, $\hat\xi$ shows absurd
patterns, e.g. the CCF plot (plot of $\hat\xi(d)$ versus $d$) shows randomly changing inflection, or $|\hat\xi(d)|$ shows an overall increasing trend against $d$ rather than decreasing, which are not physically interpretable. This can happen when the data gets contaminated with noise and hence does not reveal the inherent useful information, or the galaxies in data sets are distributed in very different ways over the survey area. In this situation, the estimated CCF becomes useless. To overcome this
problem, we perform KPCA \citep{sch02} on the data sets (see, Section $3.5$). Then, we
consider the KPCs as our study parameters, which are used to compute $\hat\xi$.

\subsection{Kernel principal component analysis (KPCA)}\label{KPCA}

Given $M$ observations $x_{k}, k=1, 2, ..., M$, on $N$
parameters, i.e. $x_{k}\in R^{N}$, KPCA is performed on a dot product space $F$ by
using a map from $R^{N}$ to $F$ defined as,
\begin{equation*}
\Phi:R^{N}\rightarrow F.
\end{equation*}
Then, for $\sum_{k=1}^{M} x_{k}=0$, we need to solve the following equation for eigenvalues
$\lambda\geq0$ and non-zero eigenvectors $V\in F$
\begin{equation*}
\lambda V=\bar{C}V,\hspace{1 mm} where \hspace{1
mm}\bar{C}=\frac{1}{M} \sum_{j=1}^{M}\Phi(x_{j}) \Phi(x_{j})^{T}.
\end{equation*}
The above-mentioned problem eventually boils down to solving the eigen value
problem for the kernel matrix $K=((K_{ij}))_{i,j=1(1)M}$ as
follows (for details, see, \citet{sch02}),
\begin{equation*}
M \lambda \alpha=K \alpha,
\end{equation*}
where $K_{ij}=<\Phi(x_{i}), \Phi(x_{j})>$ ($<\cdot,\cdot>$
represents the usual dot product) is a positive definite (pd)
kernel. Let $\lambda_{1}\geq\lambda_{2}\geq...\geq\lambda_{M}(\geq 0)$
denote the eigenvalues of $K$ and $\alpha^{1},\alpha^{2},...,\alpha^{M}$ be the
corresponding complete set of eigenvectors, where $\lambda_{l}$
is the last nonzero eigenvalue. Then, the $k-$th KPC corresponding to $\Phi(x)$ is given by
\begin{align}\label{KPC}
\begin{split}
<V^{k},\Phi(x)>&=\sum_{i=1}^{M} \alpha^{k}_{i}<\Phi(x_{i}),\Phi(x)>\\
&=\sum_{i=1}^{M} \alpha^{k}_{i} k(x_{i},x),\hspace{1 mm}k=1,2,...,l,
\end{split}
\end{align}
where $k(x_{i},x)$= kernel corresponding to $x_{i}$ and $x$.\\
Now, the assumption $\sum_{k=1}^{M} x_{k}=0$ can be relaxed by
using the kernel matrix $\tilde{K}=((\tilde{K}_{ij}))_{i,j=1(1)M}$ in the place of $K$, where
$\tilde{K}_{ij}=(K-1_{M}K-K1_{M}+1_{M}K1_{M})_{ij},
(1_{M})_{ij}=M^{-1},$ for $i,j=1,2,...,M.$ Then, KPCA can be performed using conditionally
positive definite (cpd) kernels, which include the `pd'
kernels \citep{hof08}.

KPCA extracts the relevant nonlinear information from the raw
data in terms of the KPCs. This method is used
for feature extraction, dimension reduction, classification,
clustering, noise reduction, pattern recognition, etc. It has been
successfully applied to supernovae clustering \citep{ish12,ish13}, image denoising \citep{ras12},
clustering of gamma-ray bursts \citep{mod17}, denoising of chaotic time series \citep{jad03}, etc. In the present context, KPCA has been used as a transformation on the original data so that the noise can be significantly reduced and the actual
information underlying the data sets can be explained in terms of the KPCs.

In our study, we use the following
symmetric and `pd' kernel \citep{mod17}
\begin{equation} \label{my kernel}
k(x,y)=\exp(- {\sum\limits_{i=1}^N
|\frac{x_{i}-y_{i}}{s_{i}}|^p}),
\end{equation}
between $x,y\in R^{N}$, where $p$ $(0<p\leq 2)$ is a tuning parameter and
$s_{i}$ $(>0)$$, i=1, 2,..., N$ are scale parameters, called hyperparameters.

The higher the order of the KPC, the less relevant information and the
more noise are supposed to be contained in that component
\citep{sch02}. Hence, noise can be significantly
reduced by taking into account only the first few KPCs and discarding
the others, which is also shown in terms of pre-images in
\citet{sch02}. So, we start with the first two KPCs
and consider up to the first four KPCs (not exceeding the total number of parameters in the original data) as our study
parameters. Here, we take $s_{i}$ as the sample standard deviation corresponding to the
$i-$th parameter, and vary $p$ in a trial-and-error method. Ultimately, we choose
the number of KPCs and the value of $p$ such that we can extract
sufficient relevant information from the data sets.

\subsection{Algorithm to estimate the CCF}

In the present study, the estimate of the CCF is computed as follows.\\
(i) We consider a pair of observed
samples $D_1$ and $D_2$ of sizes $n_1$ and $n_2$ respectively. Here both samples have four parameters mass, size, $RA_l$ and $DEC_l$.\\
(ii) By KPCA, the first few (say, $N^\prime$) KPCs are extracted, using kernel \eqref{my kernel} with $s_{i}=$ sample standard deviation corresponding to the
$i-$th parameter and a specified value of $p$ in $(0,2]$, from $D_1$ and
$D_2$, denoted by $D_1^\prime$ and $D_2^\prime$
respectively. Now, our pair of study samples are $D_1^\prime$ and $D_2^\prime$ having the KPCs as study parameters.\\
(iii) We compute pair count $D_1^\prime D_2^\prime(d)$, where $d$ is the normalized
Euclidean distance computed between each of the unique pairs of galaxies in two samples.\\
(iv) For each parameter (i.e. KPC), a random sample with replacement of
size $n_1$ is drawn from the set of values of the parameter in $D_1^\prime$. Then by assigning the
parameter-wise random samples in random vectors, we form $n_1$ unclustered samples
from $D_1^\prime$, denoted by $R_1$.
Similarly, we obtain $n_2$ unclustered samples from $D_2^\prime$, denoted by $R_2$.\\
(v) We compute the pair counts $D_1^\prime R_2(d)$, $D_2^\prime R_1(d)$ and $R_1R_2(d)$.\\
(vi) We repeat steps (iv)-(v) $100$ times, i.e. draw $R_1$,
$R_2$ and compute the pair counts $100$ times. Then, the final values for
$D_1^\prime R_2(d)$, $D_2^\prime R_1(d)$ and $R_1R_2(d)$ are obtained by taking the mean of the respective pair counts.\\
(vii) Compute $\hat\xi$ using formula \eqref{ECCF}, where $D_1$ and $D_2$
are replaced by $D_1^\prime$ and $D_2^\prime$ respectively.\\
(viii) Now, we compute $\hat\xi$ for $100$ times and the mean of them is taken as the ultimate value of $\hat\xi$ and their standard error is taken as the error estimate of $\hat\xi$.\\
(ix) In step (ii), consider $N^\prime=2,3,4$ and run steps (iii)-(viii).\\
(x) Repeat step (ii) for different values of $p$, and  run steps (iii)-(ix).\\
(xi) Steps (i)-(x) are repeated considering all unique pairs of observed
samples $D_1$ and $D_2$, where $D_1$ corresponds to a data set from $1-3$ and $D_2$ corresponds to a data set from $4-8$.\\

\section{Results and discussion}

\begin{figure}
\centering
\includegraphics[width=1\textwidth]{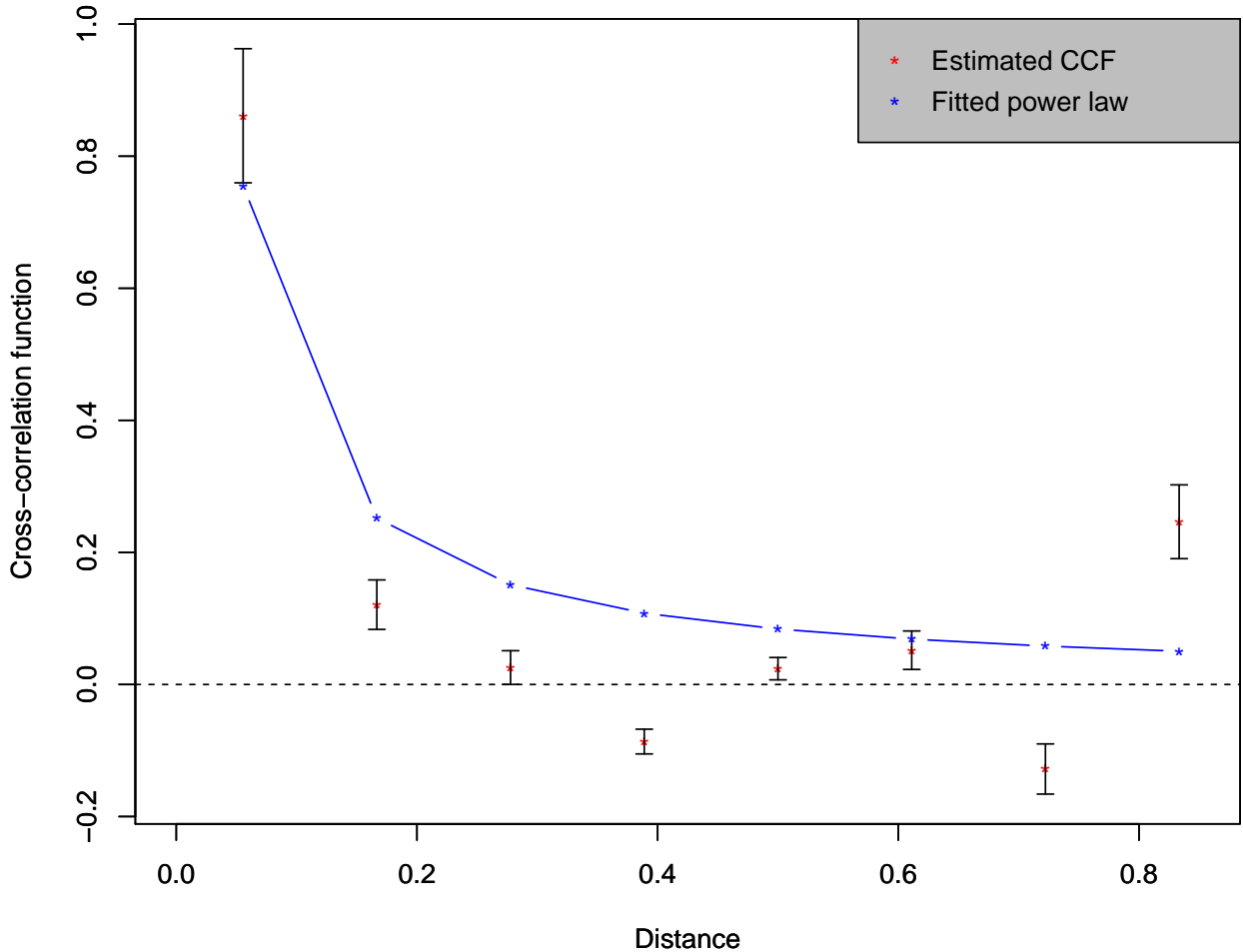}
\caption{Cross-correlation function (CCF) plot for the representative sample of the innermost part of nearby ETGs (data set $1$)
and the ETGs with $2\leq z\leq2.7$ (data set $8$). The solid line represents the fitted power law $\xi(d)=0.042d^{-1}$ (i.e. the estimated value of $A$ is $0.042)$ and $p-$value$=0.283$, corresponding to the Kolmogorov-Smirnov test for goodness of fit, indicates the fit is good.}\label{logRe18}
\end{figure}

\begin{figure}
\centering
\includegraphics[width=1\textwidth]{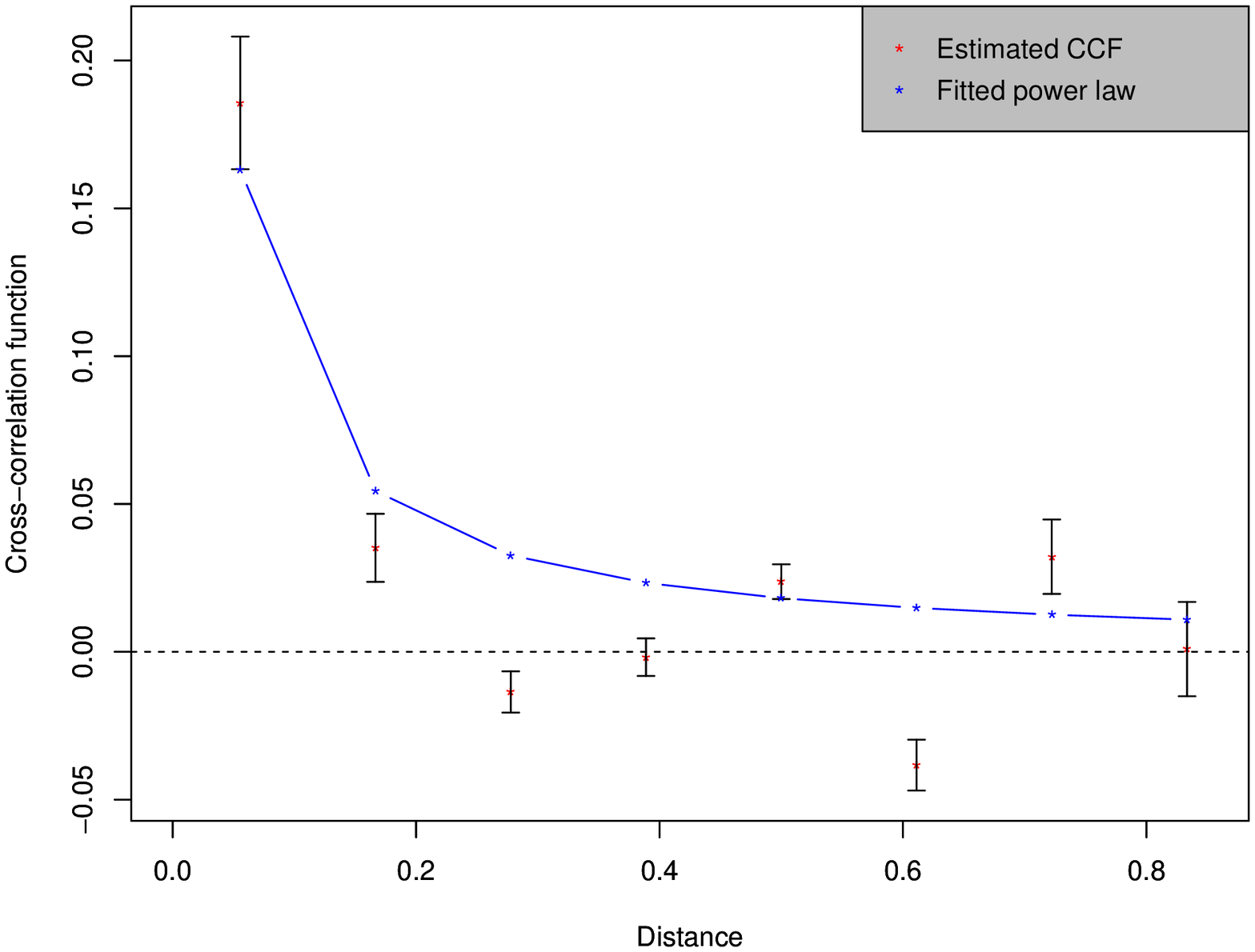}
\caption{CCF plot for the representative sample of the intermediate part of nearby ETGs (data set $2$)
and the ETGs with $0.5\leq z\leq0.75$ (data set $4$). The solid line represents the fitted power law $\xi(d)=0.009d^{-1}$ and $p-$value$=0.283$, corresponding to the Kolmogorov-Smirnov test for goodness of fit, indicates the fit is good.}\label{logRe24}
\end{figure}

\begin{figure}
\centering
\includegraphics[width=1\textwidth]{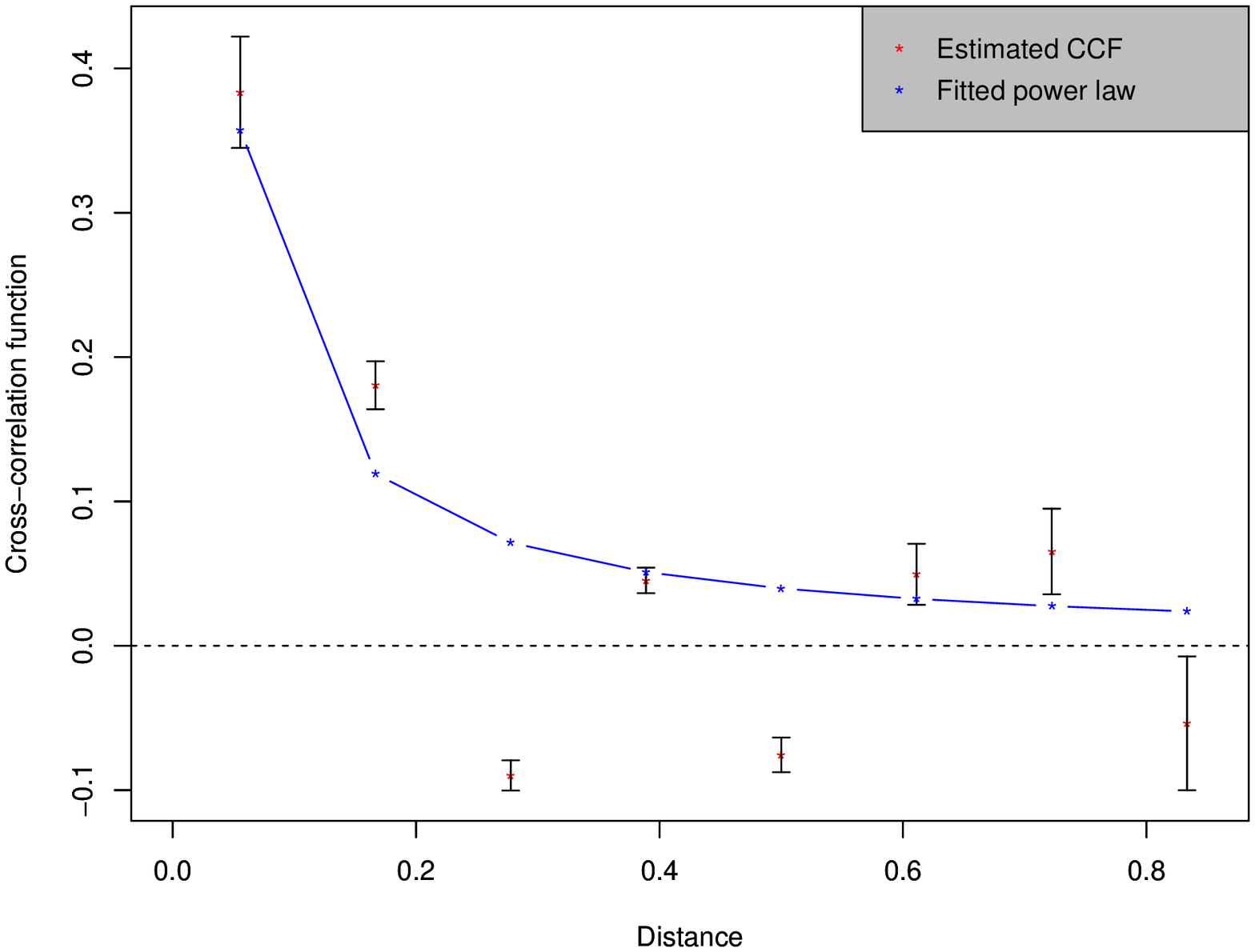}
\caption{CCF plot for the representative sample of the outermost part of nearby ETGs (data set $3$)
and the ETGs with $0.5\leq z\leq0.75$ (data set $4$). The solid line represents the fitted power law $\xi(d)=0.02d^{-1}$ and $p-$value$=0.66$, corresponding to the Kolmogorov-Smirnov test for goodness of fit, indicates the fit is quite good.}\label{logRe34}
\end{figure}

We have found
significant correlation between data set $1$ and data set $8$
(Fig. \ref{logRe18}), based on the first four KPCs
extracted from the data sets using kernel \eqref{my kernel}
with $p=0.1$ ($s_{i}=$ sample standard deviation corresponding to the
$i-$th parameter). Moderate correlation is found between data set $2$
and data set $4$ (Fig. \ref{logRe24}), using the first four KPCs
extracted through \eqref{my kernel} with $p=0.5$ from both the
data sets. Study with the first four KPCs, extracted through
\eqref{my kernel} with $p=1.5$ and $p=0.2$ from data set $3$ and
data set $4$ respectively, shows moderate correlation
(Fig. \ref{logRe34}). The above results indicate that the innermost components of massive ellipticals are strongly correlated with
ETGs in the highest redshift range, $2.0< z\leq2.7$, known
as `red nuggets', whereas both the `intermediate' and the `outermost
part' are correlated with ETGs in the redshift range $0.5<
z\leq 0.75$. It indicates two phase formation model of nearby
massive ETGs as suggested by various authors \citep{kho06,del07,guo08,kor09,hop10,naa13}.

We fit the following power law to the estimated CCF
\begin{equation}\label{pl}
\xi(d)\varpropto \frac{1}{d},\hspace{.1 in} i.e.\hspace{.1 in} \xi(d)=Ad^{-1},
\end{equation}
where $A$ is an unknown constant.
It gives us an idea of the original model, indicating correlation with $|A|>0$ and no-correlation situation with $A=0$ \citep{bla06}.
We also carry out
the Kolmogorov-Smirnov test \citep{smi48} for testing $H_0$: power law \eqref{pl} fits the estimated CCF versus $H_1$: $H_0$ is not true. The estimated value of $A$ and $p-$value of the above test are given in the CCF plots (see, Figs. \ref{logRe18}-\ref{logRe34}).

Discovery of `red nuggets' in high
redshift zone has challenged the formation of massive ellipticals through a
single event like monolithic collapse or major merger.
Nowadays ellipticals with maximum masses are expected to undergo a
number of steps rather than a single event. They are formed at $z\gtrsim 6$ through a
dissipative process, and subsequently become very massive
$\sim(10^{11} M_\odot)$ and compact $(R_e\sim 1$ kpc) in a very
short interval of time $z\sim 2$ \citep{dek09,ose10,ose12}. In spite of the above event, a
significant fraction remains less active at $z\sim2$. They are
$4-5$ times less compact and $2$ times less massive than their low-redshift descendants \citep{bui08,van08,cim08,bez09,van10,van11,whi12}.

For the massive ellipticals in the present sample, the innermost
cores (data set $1$) are strongly correlated with galaxies in the
highest redshift zone $(2.0< z\leq 2.7)$, and their respective median values for mass are approximately $10^{10} M_\odot$ and $10^{11}
M_\odot$. Hence we can reasonably conclude that this
high redshift population forms the cores of at least some, if not
all \citep{gra15,wel16}, of the present-day
massive ellipticals. Thus massive ellipticals cannot be formed by
only monolithic collapse, otherwise they would be too small and too
red \citep{van08,fer12}.

The formation of the intermediate (data set $2$) and the outer part (data set $3$)
might be explained as a result of major or minor mergers
\citep{jeo07,naa09,kav11,sha12,sha17}. Let $M_j,r_j,E_j,<v_j^2>,g_j$ and $r_{g,j}$ be respectively the mass,
radius, energy, mean squared speed, density and gravitational radius of a stellar system, where $j=i,m,f$ denote three different stages of the system as the initial stage, the stage after a merger with other systems and the final stage respectively. Then
\begin{equation*}
\frac{<v_f^2>}{<v_i^2>}=\frac{(1+\eta\epsilon)}{1+\eta}, \frac{r_{g,f}}{r_{g,i}}=\frac{(1+\eta)^2}{1+\eta\epsilon}\hspace{.03 in} and \hspace{.03 in} \frac{g_f}{g_i}=\frac{(1+\eta\epsilon)^3}{(1+\eta)^5},
\end{equation*}
where $\eta=M_m/M_i, \epsilon=<v_m^2>/<v_i^2>$. So,
$\eta=1$ implies a major merger where the size doubles. In our case, the intermediate part (data set $2$)
has mean radius $<R_{e,2}>\sim2.276$ kpc which is almost $7.689$
times larger than the mean radius of the innermost part
$<R_{e,1}>\sim0.296$ kpc. Hence we can conclude that the
intermediate parts of nearby massive ellipticals have been formed
by major mergers. The result is also consistent with previous
works \citep{cha09,cha13}.

In the limiting case, when $<v_m^2>$ $\ll$ $<v_i^2>$ or $\epsilon\ll1$,
the size increases by a factor of $4$, which implies
minor merger. In our situation, $<R_{e,3}>\sim19.554$ kpc $>>$
$<R_{e,1}>\sim0.296$ kpc. Also, the median of mass for the outer part is of
order $10^{11}M_{\odot}$, which is comparable with the
combined masses of few dwarf galaxies. This indicates that the
outermost part of nearby massive ETGs might have been formed by
several minor mergers. The combined study of intrinsic parameters with spatial effects gives a
more robust picture for the two-phase formation of massive
ellipticals and overcomes the conflict of three phase formation \citep{de14}. Our result is also consistent with that
of \citet{mon08}, where halos of massive
galaxies are found to be formed by tidal stripping of satellite
dwarf galaxies.

\section{Acknowledgments}
The authors would like to thank the Editor for helpful feedback and the anonymous referee for constructive comments to improve the quality of the paper. Tanuka Chattopadhyay and Asis Kumar Chattopadhyay had been partially supported
by Indo-French project (Project No. 15EP06, 2015-17) for the
work.

\clearpage

\clearpage

\clearpage

\clearpage

\clearpage

\clearpage

\clearpage

\clearpage

\clearpage

\clearpage
\begin{table}
\small \caption{Compatibility test between two data sets in the common redshift zone, using the parameters size, mass, RA$_l$ and DEC$_l$.}
\begin{center}
\begin{tabular}{c c c c c c}
\tableline\tableline
\noalign{\vskip .1in}
First  & Second  & $z$ & No. of ETGs   & Parameter-wise & Decision at 5\% \\
data set & data set  &     & in two data sets & $p-$values &  level of significance    \\
\noalign{\vskip .1in}

\tableline
\noalign{\vskip .1in}
Damjanov        & Papovich  & [1.49, 1.79] & (22, 248) & (1, 1, 0.413, 1) & Accept\\
et al. $(2011)$ & et al. $(2012)$& & & &\\
\noalign{\vskip .3in}
              & Chen     & [0.703, 1.309] & (199, 142) &(1, 0.998, 1, 1) & Accept\\
''                & et al. $(2013)$ &&&&\\
                \noalign{\vskip .3in}
           & Szomoru     & (0.5,2.7) &  (364, 177) &(1, 1, 1, 1) & Accept\\
''                   & et al. $(2013)$ &&&&\\
                \noalign{\vskip .3in}
              & McLure     & [1.280, 1.505] & (36, 81) &(1, 1, 0.389, 1) & Accept\\
''                & et al. $(2013)$  &&&&\\
                \noalign{\vskip .1in}
\tableline
\end{tabular}
\end{center}
\label{table:t1}
\end{table}

\end{document}